\documentstyle[aps,pra,epsfig]{revtex}

\begin{document}

\draft

\title{Temporal imperfections building up
correcting codes}

\author{Stefano Mancini\footnote{Present address:
Dipartimento di Fisica, Universit\`a di Camerino,
I-62032 Camerino, Italy}   and 
Rodolfo Bonifacio}

\address{
INFM, Dipartimento di Fisica,
Universit\`a di Milano,
Via Celoria 16, I-20133 Milano, Italy
}

\maketitle

\widetext

\begin{abstract}
We address the timing problem in realizing 
correcting codes for quantum information processing.
To deal with temporal uncertainties we employ a consistent
quantum mechanical approach.
The conditions for optimizing the effect of error correction
in such a case are determined.  
\end{abstract}

\section{Introduction}

Quantum error correction protects quantum information against 
environmental noise \cite{NC00}.
After the initial discovery of quantum error correction codes 
\cite{SHOR95,STEANE96} significant progress has been made in the 
development and understanding of these codes.
Of particular significance has been the discovery of 
minimal codes \cite{BEN96}, which are the simplest codes 
able to correct amplitude and/or phase errors \cite{NC00}.
Also the possibility of encoding and decoding 
in presence of noise has been demonstrated \cite{FAULT}.
However, these fault-tolerant schemes are extremely complicated and 
involve many more qubits than simple codes.
Their experimental implementation is therefore unlikely,
at least in a near future.

On the other hand, the issue of errors arising during encoding
and decoding has been partially investigated in the simplest
error correcting codes \cite{EDNOISE}.
These are the codes that would, hopefully, be implemented 
in a near future.
Along this line, we would address the question 
of how timing problems affect the performances of 
simple codes.

As matter of fact, encoding and decoding procedures, 
even for simple codes, take place in several steps
requiring turning on and off given interactions.
Hence, it should be natural to deal with time uncertainties 
leading to noisy effect.
These may be due, for instance,  
to the timing of laser pulses 
\cite{WINE}, or RF fields \cite{NMR}, 
and might compromise the correction procedure.

Thus, the aim of this work is to study the circumstances
under which a simple correcting code results beneficial 
notwithstanding temporal imperfections during
encoding and decoding. To this end we shall exploit a
quantum mechanical consistent approach \cite{RB}.

\section{Perfect error correction}

Let us consider a single qubit
in a two dimensional Hilbert space.
A convenient basis is given by the eigenstates of the
Pauli matrix $\sigma_z$. We denote them as
$\{|0\rangle\,,\,|1\rangle\}$.
Then, the environmental effects on the qubit can be 
described by means of the Lindblad master equation
(in natural units)
\begin{equation}\label{ME}
\dot\rho=-i\left[H,\rho\right]
+\sum_j\left(L_j\rho L_j^{\dag}
-\frac{1}{2}L_j^{\dag}L_j\rho
-\frac{1}{2}\rho L_j^{\dag}L_j\right)\,,
\end{equation}
where $H$ is the system Hamiltonian and
$L_j$ are the Lindblad operators representing the
interaction with the environment.
We are now going to consider the free Hamiltonian $H=\sigma_z$
having unit frequency.
Furthermore, the most general interaction is represented by the 
isotropic noise with Lindblad operators 
$L_1=\sqrt{\gamma}\sigma_x$, 
$L_2=\sqrt{\gamma}\sigma_y$, 
$L_3=\sqrt{\gamma}\sigma_z$; $\gamma$ being the 
decoherence rate.

For a generic initial state 
\begin{equation}\label{PSI}
|\psi\rangle=c_0|0\rangle+c_1|1\rangle\,,
\end{equation} 
the solution of the master equation (\ref{ME}), 
in case of isotropic noise, reads
\begin{eqnarray}\label{RHOT}
\rho(t)&=&
\left[\frac{|c_0|^2}{2}\left(1+e^{-4\gamma t}\right)
+\frac{|c_1|^2}{2}\left(1-e^{-4\gamma t}\right)\right]
|0\rangle\langle 0|
\nonumber\\
&+&
\left[\frac{|c_0|^2}{2}\left(1-e^{-4\gamma t}\right)
+\frac{|c_1|^2}{2}\left(1+e^{-4\gamma t}\right)\right]
|1\rangle\langle 1|
\nonumber\\
&+&\left[\frac{c_0c_1^*}{2}\left(e^{-2\gamma t}
+e^{-6\gamma t}\right)\right]
|0\rangle\langle 1|e^{2it}
\nonumber\\
&+&\left[\frac{c_0^*c_1}{2}\left(e^{-2\gamma t}
+e^{-6\gamma t}\right)\right]
|1\rangle\langle 0|e^{-2it}\,.
\end{eqnarray}

In the following we shall consider a simple information process,
i.e. the information storage. Thus the single qubit dynamics
is exactly described by Eq.(\ref{ME}).
The probability that the qubit (\ref{PSI}) remains error free  
(for isotropic noise) after a time $t$, is given by
\begin{equation}\label{PSDEF}
{\cal P}_s(t)={\rm Tr}\left\{\rho(t)\rho_{rev}(t)\right\}_{ave}\,,
\end{equation}
where the subscript $rev$ means the evolution under the reversible 
part of the master equation (i.e. only that containing $H$), while 
$ave$ means the average overall possible
states (all possible values of $c_0$ and $c_1$).
It results, for a storage time $T$, that
\begin{equation}\label{PSEXP}
{\cal P}_s(T)=\frac{1}{2}+\frac{1}{4}e^{-4\gamma T}
+\frac{1}{8}\left(e^{-2\gamma T}+e^{-6\gamma T}\right)\,.
\end{equation}

Consider now the $5$-qubit encoding and decoding procedure 
\cite{BEN96} which 
is able to correct perfectly for a single error in one of the $5$ 
qubits, but fails if there are two or more errors 
\footnote{To be precise, this code can also correct 
some double errors.}. 
Then, the probability of survival of a single encoded qubit state
for time $T$ is the sum of the zero error and one error 
probabilities; that is
\begin{equation}\label{PSS}
{\cal P}^*_s(T)=[{\cal P}_s(T)]^5+5[{\cal P}_s(T)]^4
\times[1-{\cal P}_s(T)]\,,
\end{equation}
where we assumed each qubit suffering the same decoherence rate.
The star superscript on $P_s$ reminds us that the probability 
refers to the encoded qubit.

\section{Imperfect error correction}

Consider now the case where encoding and decoding procedures
are not immune from the noise. Specifically, we wish to consider 
the case where only timing problems occur; so, we assume
the isotropic noise to be negligible during this stages.
This could be reasonable if the decoherence rate is very small
and the encoding (decoding) time is much smaller than the
storage time. Then, we denote with $T_{ed}/2$ the time for the 
encoding procedure. The same is also true for the reverse 
process, the decoding. So that the total encoding+decoding time
would be $T_{ed}$ with $T_{ed}\ll T$. 

To account for timing problems we exploit a recent theory
developed by one of us \cite{RB}.
Namely,
the evolution of a system is averaged on 
a suitable probability 
distribution $\wp(t,t')$ where $t'$ represents all possible 
times within the ensemble.
Let $\rho(0)$ be the initial state, then
the evolved state would be 
\begin{equation}\label{RHOBAR}
{\overline\rho}(t)=\int^{\infty}_0\, dt' \, \wp(t,t')\, \rho(t')\,,
\end{equation}
where $\rho(t')=\exp\{-i{\cal L}t'\}\rho(0)$ 
is the solution of the 
Liouville-Von Neumann equation.

One can write as well 
\begin{equation}\label{RHOBARU}
{\overline\rho}(t)={\cal U}(t)\rho(0)\,, 
\end{equation} 
where the superoperator ${\cal U}$ is given by
\begin{equation}\label{UP} 
{\cal U}(t)=\int^{\infty}_0\, dt' \, \wp(t,t')\,e^{-i{\cal L}t'}\,.  
\end{equation} 
In Ref. \cite{RB}, the
function $\wp(t,t')$ has been determined to satisfy 
the following conditions:
i) ${\overline\rho}(t)$
must be a density operator; ii) ${\cal U}(t)$ satisfies
the semigroup property. 
These requirements are satisfied by 
\begin{equation}\label{U} 
{\cal U}(t)=\frac{1}{(1+i{\cal L}\tau)^{t/\tau}}\,,  
\end{equation} 
and  
\begin{equation}\label{WP} 
\wp(t,t')=\frac{1}{\tau}\frac{e^{-t'/\tau}}{\Gamma(t/\tau)}
\left(\frac{t'}{\tau}\right)^{(t/\tau)-1}\,, 
\end{equation} 
where $\Gamma(.)$ is the Gamma function and
the parameter $\tau$ naturally appears as a scaling time.
Its meaning can be
understood by considering the mean $\langle t'\rangle=t$, 
and the variance
$\langle t'^2\rangle-\langle t'\rangle^2=\tau t$. 
Hence, $\tau$ rules the strength of time fluctuations,
or, otherwise, the characteristic correlation time of 
fluctuations. 
When $\tau\to 0$, $\wp(t,t')\to \delta(t-t')$ so
that ${\overline\rho}(t)\equiv\rho(t)$ 
and ${\cal U}(t)=\exp\{-i{\cal L}t\}$ 
is the usual evolution.

Coming back to our problem, 
we consider the reversible evolution in Eq.(\ref{RHOT}) 
(i.e., $\gamma\to 0$),
and we then average obtaining $\overline{\rho}(t)$
accordingly to Eq.(\ref{RHOBAR}).
Therefore, the probability that the system remains
unaffected by timing errors is
\begin{equation}\label{PT}
{\cal P}_t(t)={\rm Tr}\left\{\overline{\rho}_{rev}(t)
\rho_{rev}(t)\right\}_{ave}\,.
\end{equation}
Roughly speaking, this can be used to calculate the probability 
of no errors per qubit during encoding+decoding procedure, 
that is
\begin{equation}\label{PTTED}
{\cal P}_t(T_{ed})=\frac{3}{4}+\frac{1}{4}{\rm Re}
\left\{(1+2i\tau)^{-T_{ed}/\tau} e^{2iT_{ed}}\right\}\,.
\end{equation}

Thus, the probability that there is no error in the $5$-qubit
system is
the product of the probability (\ref{PTTED}) of all $5$ qubits 
surviving $T_{ed}$ with the probability (\ref{PSS}) of zero or one
error (which can be corrected) during the time $T$. This leads to
\begin{equation}\label{PCAL}
P(T_{ed},T)=[{\cal P}_t(T_{ed})]^5\times {\cal P}_s^*(T)\,.
\end{equation}
We now introduce a parameter which markers the efficiency
of the correction procedure. Namely, we introduce the 
ratio of the mismatch without correction to the mismatch with
correction
\begin{equation}\label{RMIS}
{\cal R}=\frac{1-[{\cal P}_s(T+T_{ed})]^5}{1-P(T_{ed},T)}\,.
\end{equation}
Provided that ${\cal R}$ stays above unit value, there should be 
benefit from error correction even though timing errors occur 
during encoding+decoding.

In Figure 1 we show the contours of ${\cal R}$ in the plane 
of parameters $\tau$ and $T_{ed}$. 
From bright to dark region the correction procedure 
becomes less efficient.
In the black zone it results useless since ${\cal R}<1$.

\begin{figure}[t]
\centerline{\epsfig{figure=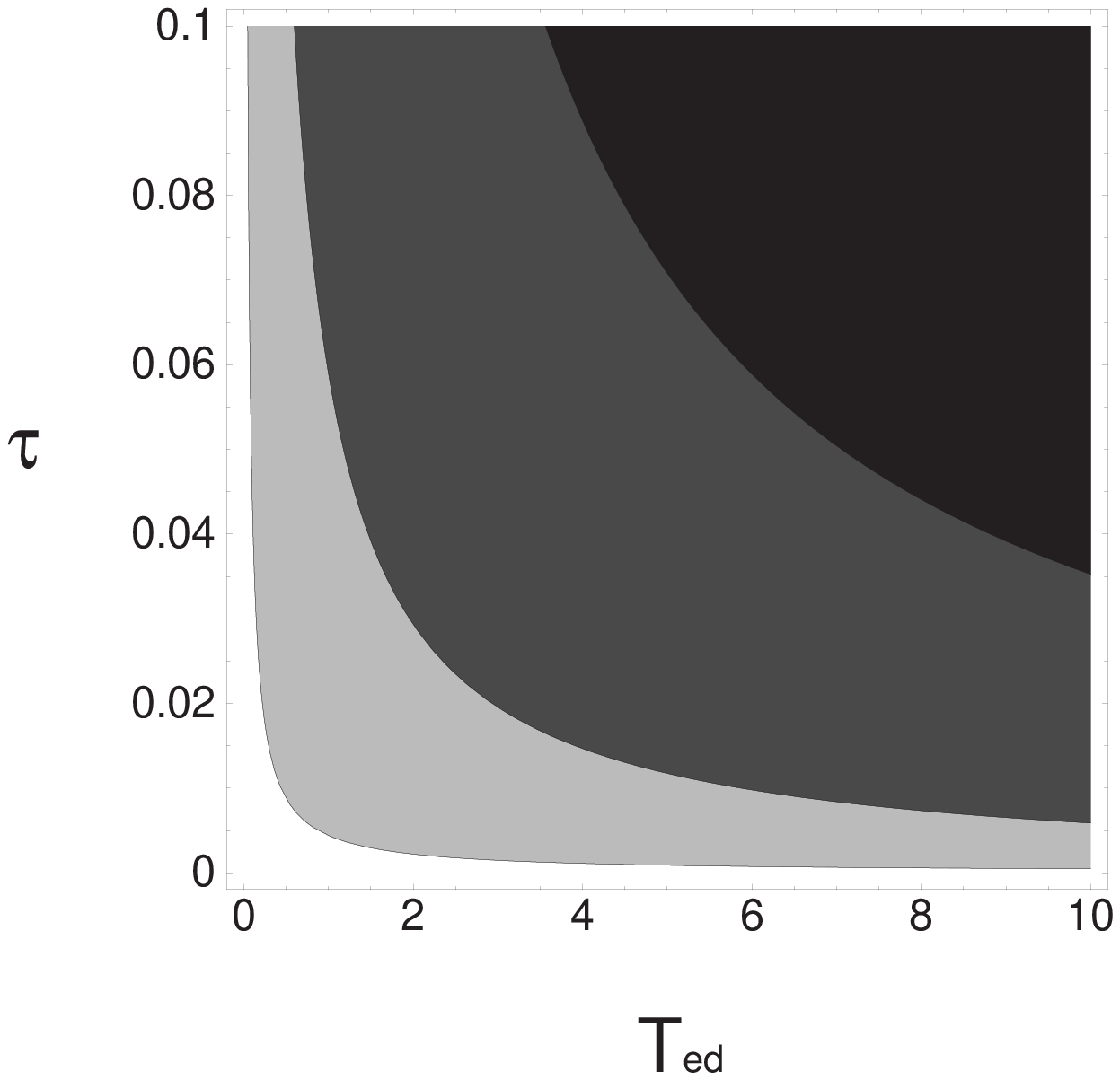,width=3.0in}}
\caption{Contour plot of the quantity ${\cal R}$ as 
function of $\tau$ and $T_{ed}$.
The values of other parameters are
$\gamma=10^{-5}$, $T=10^4$ (frequencies and times are
dimensionless since they are properly scaled through the 
system frequency).
Contours values are $3$, $2$ and $1$,
moving from bright to dark region.
}
\label{fig1}
\end{figure}

\section{Repeated correction procedure}

The number of correction procedures applied during the time 
$T$ can be varied.
Consider the problem of optimizing error correction to achieve the 
greatest probability success for the storage of a qubit state, given
the freedom to apply an arbitrary number $N$ of 
encoding+decoding procedures during $T$. 
Assume that these are spaced out equally. In case of perfect
error correction, it is obviously beneficial to apply as many
corrections as possible. The probability of success for $N$ 
applications is
\begin{equation}\label{PN}
{\cal P}_N=[{\cal P}_s^*(T/N)]^N\,.
\end{equation}
This maximizes for $N\to\infty$, tending to unity. Such behavior
is like the Zeno or watchdog effect; there is no change at all from
the initial state as $N\to\infty$ \cite{ZENO}.

However, in case of timing errors, there should be an optimum 
value of $N$. The generalization of Eq. (\ref{PCAL}) 
to $N$ equally spaced corrections is
\begin{equation}
P_N=[{\cal P}_t(T_{ed})]^{5N}\times[{\cal P}_s^*(T/N)]^N\,.
\end{equation}

Then, in Figure 2 we plot $P_N$ as function
of $N$. We see that the optimum number $N$ which maximize
the probability decreases by increasing the value of $\tau$.
In figure, we pass from $N_{opt}=6$ for $\tau=0.003$,
to $N_{opt}=3$ for $\tau=0.01$, and $N_{opt}=1$ for $\tau=0.05$.
Beyond that value the error correction becomes useless.

\begin{figure}[t]
\centerline{\epsfig{figure=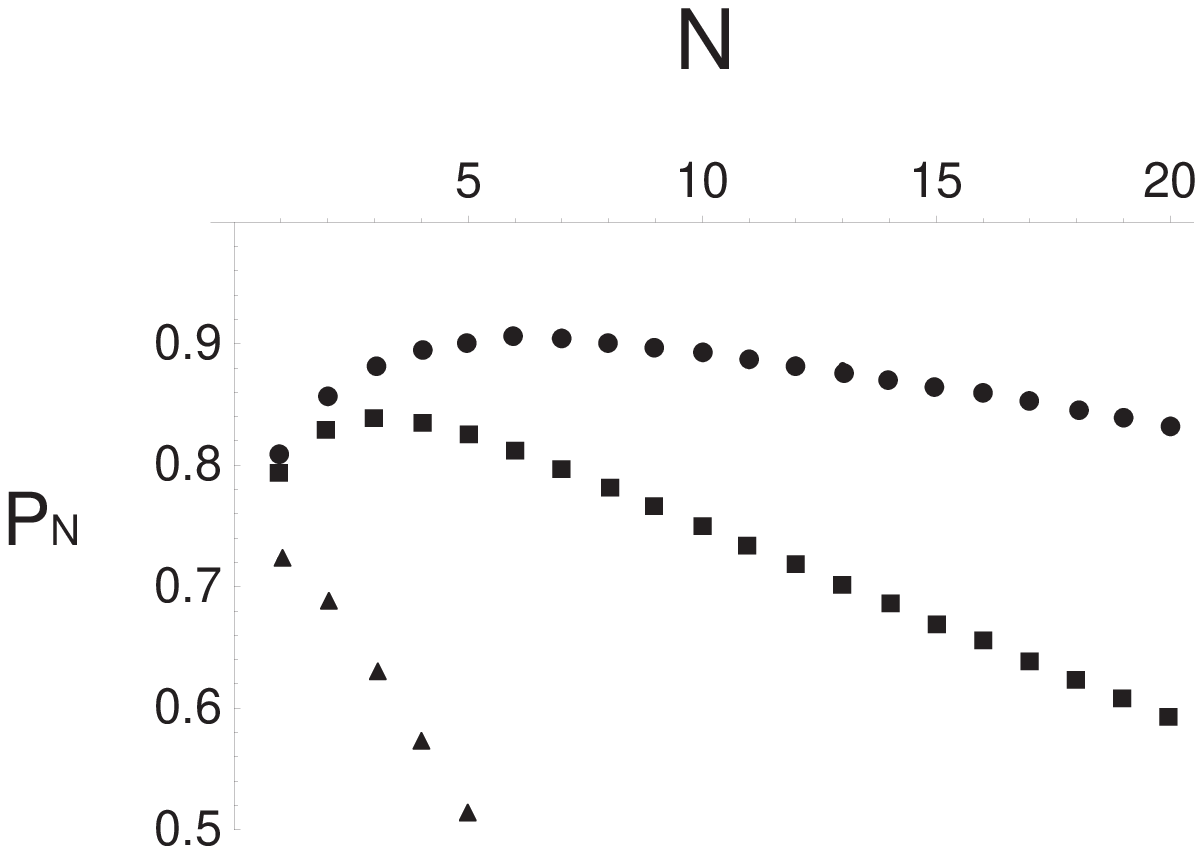,width=3.0in}}
\caption{Success probability $P_N$ 
for $N$-repeated correction
procedures as function of $N$, for different values of
$\tau$ (circles $\tau=0.003$; squares $\tau=0.01$; 
triangles $\tau=0.05$).
The values of other parameters in frequency units are
$\gamma=10^{-5}$, $T=10^4$, $T_{ed}=1$
(frequencies and times are
dimensionless since they are properly scaled trough the 
system frequency).
}
\label{fig2}
\end{figure}

\section{Conclusion}

In conclusion we have considered the problem of 
temporal imperfections in building up correcting 
codes\footnote{Whenever timing
errors become very small one could even
employ fault tolerant correction procedure (treating them
as phase errors), but this is not within reach.}. 
To this end we have exploited a quite ductile model based on 
random time evolution.
 
In Ref. \cite{NDD} non-dissipative decoherence bounds 
for ion-trap based quantum computation were established
by estimating $\tau\approx 10^{-3}$.
In such a case $\tau$ gives an estimate of the pulse are 
fluctuations for the laser inducing transitions. 
This value of $\tau$ gives
the restriction $T_{ed}/T<10^{-3}$ for the success 
of the above considered code.

Finally, we recognize that
our approach is somewhat rough since 
the reversible dynamics during encoding
and decoding has been identified with
the free dynamics. Nevertheless, it allows 
an estimation of realistic performances of 
simple challenging codes. 
A more accurate study is left for future work.

\section*{Acknowledgments}
We are indebted to David Vitali for his useful comments.

\end{document}